\documentstyle[multicol,aps,pre,epsf]{revtex}
\begin{document}
\title{Spatial organization in cyclic Lotka-Volterra systems}
\author{L.~Frachebourg$^\ast$, P.~L.~Krapivsky$^{\ast}\dag$, 
and E.~Ben-Naim$\ddag$} 
\address{$^\ast$Center for Polymer Studies and Department of Physics,
Boston University, Boston, MA 02215}
\address{$\dag$Courant Institute of Mathematical Sciences, 
New York University, New York, NY 10012-1185}
\address{$\ddag$The James Franck Institute, The University of Chicago, 
Chicago, IL 60637}
\maketitle
\begin{abstract}
We study the evolution of a system of $N$ interacting species which
mimics the dynamics of a cyclic food chain.  On a one-dimensional
lattice with $N<5$ species, spatial inhomogeneities develop
spontaneously in initially homogeneous systems.  The arising spatial
patterns form a mosaic of single-species domains with algebraically
growing size, $\ell(t)\sim t^\alpha$, where $\alpha=3/4$ (1/2) and 1/3
for $N=3$ with sequential (parallel) dynamics and $N=4$, respectively.
The domain distribution also exhibits a self-similar spatial structure
which is characterized by an additional length scale, ${\cal L}(t)\sim
t^\beta$, with $\beta=1$ and 2/3 for $N=3$ and 4, respectively.  For
$N\geq 5$, the system quickly reaches a frozen state with non
interacting neighboring species.  We investigate the time distribution
of the number of mutations of a site using scaling arguments as well as
an exact solution for $N=3$.  Some possible extensions of the system are
analyzed.

{PACS numbers:  02.50.Ga, 05.70.Ln, 05.40.+j}
\end{abstract}
\begin{multicols}{2}

\section{Introduction}

The classic Lotka-Volterra equations\cite{lotka,volterra,montroll}
mimic the dynamics of interacting species such as predator-prey
systems.  These equations are rather successful in predicting density
oscillations which are known to exist in Nature.  For spatially
inhomogeneous situations, Lotka-Volterra equations\cite{murray} are
straightforwardly generalized to diffusion-reaction
equations\cite{pattern}; these equations were widely applied to more
complex ecological processes.  However, such an approach ignores
spatial correlations and therefore fails to predict the development of
spatial heterogeneities in initially homogeneous systems.  For
chemical processes, the crucial role that spatial heterogeneities play
in governing the kinetics has been appreciated over the past decade,
see {\it e.g.}\cite{redner} and references therein.  Therefore, in low
spatial dimensions the mean-field like rate equations approach (analog
of the Lotka-Volterra equations in chemical kinetics) fails to provide
the correct asymptotic behavior.  Indeed, a homogeneous initial state
evolves to a strongly heterogeneous state, namely to a coarsening
mosaic of reactants which confines the actual microscopic reaction to
the interfacial regions between domains, and therefore the kinetics
are significantly slowed down.  Similar spatial organization was
recently reported in Lotka-Volterra systems\cite{t,ti,tome,may,sole}.
However, theoretical understanding of these systems is still
incomplete.

In this study, we consider the evolution of an $N$ species food chain,
where every species plays the role of prey and predator
simultaneously.  The food chain is thus assumed to be cyclic; {\it
e.g.}, in the 3-species system, $A$ eats $B$, $B$ eats $C$, and $C$
eats $A$.  Every ``eating'' event leads to duplication of the winner
and elimination of the loser, therefore the 3-species food chain is
symbolized by the reaction scheme
\begin{equation}    
A+B\to 2A, \quad
B+C\to 2B, \quad
C+A\to 2C.  
\label{ABCscheme}
\end{equation}
The corresponding stochastic process is well defined on a lattice,
where the interaction is restricted to nearest neighbor sites.
Initially, every lattice site is assumed to be occupied; clearly, the
lattice then remains fully occupied.

Given the simplicity of the reaction process (\ref{ABCscheme}), one
anticipates that it can provide a caricature description of a number
of phenomena in Nature and Society.  One example is the voter
model\cite{lig,elp}, which is applicable to chemical reactions on
catalytic surfaces\cite{zgb,lp}.  This model is equivalent to the
2-species model, which is described by the reaction scheme $A+B\to 2A$
or $A+B\to 2B$ (both channels are equally probable).  The subsequent
cyclic $N$-species generalization is also called the $N$-color cyclic
voter model\cite{bramson}.

The rest of this paper is organized as follows. In section II, we
discuss predator-prey dynamics in infinite dimensions (mean-field) and
a relationship to Nambu mechanics. Section III examines interface
dynamics in one dimension. We analyze the corresponding rate equations
and show that spatial organization into an alternating mosaic of
growing domains occurs for $N<5$ only.  While the qualitative
predictions made in Sec. III are correct, the quantitative predictions
fail. In Sec. IV, we further analyze the interface dynamics using
primarily scaling arguments and numerical simulations for the most
interesting cases, $N=3$ and 4. We consider both sequential and
parallel dynamics evolution rules, as the system may be sensitive to
such rules. Section V studies the dynamics of mutations and 
quantities such as the fraction of
persistent sites. In sections III-V we focus on symmetric and
uncorrelated initial conditions where all the $N$ species have an
initial density of $1/N$.  In section VI, we describe several natural
generalizations of the model to asymmetric initial concentrations,
symmetric interaction rules and reaction-diffusion descriptions.
Section VII discusses our results within the general framework of
coarsening phenomena. A Summary is presented in section VIII.

\section{DYNAMICS IN INFINITE DIMENSION}

Since we are interested in the role of spatial correlations, we study
primarily the extreme case of one dimension, where spatial
inhomogeneities are most pronounced.  As a preliminary step, however,
it proves useful to examine the opposite extreme where no correlations
are present, {\it i.e.}, the cyclic Lotka-Volterra system on a
complete graph.  On this structure, all sites are neighbors and the
spatial structure is irrelevant, {\it i.e.}, $d=\infty$.

Consider the 3-species Lotka-Volterra model (\ref{ABCscheme}).  The
simplest mean-field description of this model, which is {\it exact}
only on a complete graph, consists of the following rate equations for
the species concentrations $a(t)$, $b(t)$, and $c(t)$,
\begin{eqnarray}  \label{abc}  
   \dot a&=&a(b-c),\nonumber\\ 
   \dot b&=&b(c-a),\\ 
   \dot c&=&c(a-b),\nonumber
\end{eqnarray}
where the over-dot denotes the time derivative.  These equations obey
the trivial conservation law, $a+b+c=1$, that merely reflects particle
conservation.  There is an additional hidden integral, $abc={\rm
const}$.  The existence of these two integrals allows us to describe the
behavior of the system without actually solving Eqs.~(\ref{abc}).
Clearly, a generic solution to Eqs.~(\ref{abc}) is periodic; this
solution can be expressed through elliptic functions.  There are
also simpler exceptional solutions, corresponding to situations
when one of the concentrations is zero.  In this case, the solution
approaches an attracting point; {\it e.g.}, if $c=0$, one readily
finds that $a(t)\to 1$ and $b(t)\to 0$ when $t\to\infty$, while the
approach towards these limiting values is exponential in time. 

Several authors notified that some Lotka-Volterra equations admit a
Hamiltonian structure\cite{hamilt}.  However, even in very simple
situations the Hamiltonian structure can be impossible, {\it e.g.}, it
is clearly the case when the number of species is {\it odd} as in the
3-species cyclic Lotka-Volterra system.  We now describe a more
intricate relationship between the 3-species Lotka-Volterra equations
and Nambu mechanics.  Nambu mechanics\cite{nambu,takht} is a
generalized Hamiltonian mechanics where the usual pair of canonical
variables of the Hamiltonian are replaced by a triplet of coordinates.
Nambu formulated these dynamics via the ternary operation, the Nambu
bracket, replacing the binary Poisson bracket.  The Nambu bracket,
$\{f,g,h\}$, satisfies the usual skew-symmetry, the Leibniz rule, and
the generalized Jacobi identity\cite{takht}.  Let $c_1,c_2,c_3$ denote
the coordinates.  The canonical Nambu bracket is
\begin{equation}    
\{f_1,f_2,f_3\}={\partial (f_1,f_2,f_3)\over \partial (c_1,c_2,c_3)},
\label{bracket}
\end{equation}
where the right-hand side is the Jacobian.  Similarly, one can define
the canonical $N$-ary Nambu bracket, 
${\partial (f_1,\ldots,f_N)\over \partial (c_1,\ldots,c_N)}$, 
which is identical to the canonical Poisson bracket for $N=2$.  To define 
the Hamiltonian dynamics one needs a special function $H$; then, an 
arbitrary function $f$ evolves according to Hamilton's equations of
motion, $\dot f=\{f,H\}$.  To define the Nambu dynamics, one needs
two ``Hamiltonians'' denoted by $H_1$ and $H_2$.  Nambu's
equations of motion read 
\begin{equation}    
\dot f=\{f,H_1,H_2\}.
\label{motion}
\end{equation}

A few systems possessing hidden integrals of motion were immersed in
the framework of Nambu mechanics\cite{nambu,takht,chak,chat}.
Surprisingly, Eqs.~(\ref{abc}) provide a very simple additional example.
Indeed, writing $a=c_1, b=c_2, c=c_3$, using the integrals of
Eqs.~(\ref{abc}) as the Hamiltonians, $H_1=a+b+c$ and $H_2=abc$, 
choosing the canonical Nambu bracket (\ref{bracket}), and specializing
Nambu equations (\ref{motion}) to $f=a,b,c$, we recover Eqs.~(\ref{abc}). 

One can investigate Lotka-Volterra models for a greater number of
species.  We checked that for $N=4$, one can still recast the
corresponding Lotka-Volterra equations for the concentrations $c_j,
1\leq j\leq 4$, into the framework of Nambu mechanics with canonical
Nambu bracket and the Hamiltonians $H_1=c_1+c_2+c_3+c_4$,
$H_2=c_1c_3$, and $H_3=c_2c_4$.  It would be interesting to establish
a Nambu-type formulation (if any) for the cyclic $N$-species model
with arbitrary $N$.  On the complete graph, the corresponding rate
equations are 
\begin{equation}    
\dot c_k=c_k\left(c_{k+1}-c_{k-1}\right)\equiv F_k, 
\label{lotka}
\end{equation}
for $k=1,\ldots,N$, and the addition and subtraction of the index are
taken modulo $N$.  Eqs.~(\ref{lotka}) appear not only in mathematical
biology\cite{volterra} but also in plasma physics where they describe
the fine structure of the spectrum of Langmuir oscillations\cite{zakh} 
(the system (\ref{lotka}) with $-\infty<k<\infty$ is also called
``Langmuir lattice'').

The possible existence of the Nambu formulation implies that
Eqs.~(\ref{lotka}) should satisfy the Liouville condition\cite{liouv},
$\sum_k \partial F_k/ \partial c_k=0$.  An elementary computation
shows that this {\it necessary} condition is indeed satisfied.
Another necessary condition is the existence of $N-1$ independent
integrals of motion.  For the general $N$-species cyclic
Lotka-Volterra model one easily finds two such integrals,
\begin{equation}    
H_1=\sum_{j=1}^N c_j, \quad{\rm and}\quad 
H_2=\prod_{j=1}^N c_j.
\label{int}
\end{equation}
When the number of species is even, $N=2P$, the products of even and odd
concentrations separately, 
\begin{equation}    
H_3=\prod_{j=1}^P c_{2j}, \quad
H_4=\prod_{j=1}^P c_{2j-1},
\label{intadd}
\end{equation}
are also conserved so there are at least three independent integrals
of motion: $H_1$, $H_3$, and $H_4$.  Generally, $[N/2]$ integrals of
Eqs.~(\ref{lotka}) involving polynomials of $c_j$ and $\ln c_j$ are
known (see {\it e.g.}\cite{bog,moser} and references therein).
However, for the $N$-species cyclic Lotka-Volterra model to admit the
Nambu formulation, $N-1$ independent integrals are necessary, and 
it is not clear whether a Nambu structure exists for $N\geq 5$.

Although we have not yet derived new results with the help of the Nambu
structure, its richness may be useful in understanding the evolution
of cyclic Lotka-Volterra systems.

\section{RATE EQUATIONS FOR 1D INTERFACE DYNAMICS}

The infinite dimensional analysis fails to describe the dynamics of the
actual stochastic process in low dimensions.  For example, while the sum
$H_1=\sum_j c_j$ is conserved, the product $H_2=\Pi_j c_j$ is not a
conserved quantity in 1D.  Furthermore, the structure of
Eqs.~(\ref{abc}) does not address fluctuations in the spatial
distribution of interacting populations.  For $N<5$, we shall see that
the spatial structure evolves forever, single-species domains arise and
grow indefinitely, and the process exhibits coarsening.  In other words,
equilibrium is never achieved and instead a networks of domains
develops.  The domain patterns are self-similar, {\it i.e.}, the
structure at later times and at earlier times differ only by a global
change of scale. Such a behavior is a signature of {\it dynamical
scaling}.

In the following section, we study the motion of ``domain walls'',
namely, interfaces separating domains of different species. For the
$N$ species process a bond connecting two sites is an interface bond
if the corresponding two sites are occupied by two different
species. Thus, there are $N-1$ independent types of interfaces, of
which $N-3$ are immobile and 2 are mobile. For symmetric initial 
conditions ($c_i(0)=1/N$), the different types of interface bonds 
are present with 
initial concentration equal to $1/N$ (with
probability $1/N$ a given bond does not contain an interface).
Interfaces move and react according to $N$-dependent rules defined
below.  For large $N$, most interfaces are immobile and the system
quickly reaches a state where all mobile interfaces are eliminated.

\subsection{2-Species}

Consider the simplest case of $N=2$, where there are two equivalent
interfaces ($AB$ and $BA$), denoted by $I$. An isolated interface
performs a random walk, {\it i.e.}, it hops to one of its nearest
neighbors.  When two interfaces meet they annihilate.  The
corresponding reaction scheme is therefore $I+I\to \emptyset$. Assuming that
neighboring interfaces are uncorrelated, we arrive at the binary reaction
equation
\begin{equation}
\dot I=-4I^2.
\label{idot}
\end{equation}
The hoping rate was taken as unity without loss of generality.
Solving this equation subject to the initial conditions $I(0)=1/2$
gives
\begin{equation}
I(t)={1\over 2+4t}.
\label{i} 
\end{equation}
The system evolves into a mosaic of alternating domains $AABBBAAABBB$.
The typical size of a domain grows linearly with time, ${\cal
L}(t)\sim t$. 

\subsection{3-species}

In the case $N=3$, there are two types of interfaces, right moving
($AB$, $BC$, and $CA$), and left moving ($BA$, $CB$, and $AC$),
denoted by $R$ and $L$, respectively.  Starting with a symmetric
initial distribution, all right (left) interfaces are equivalent.  When a
right moving interface meets a left moving one, they annihilate
$R+L\to \emptyset$. When a right moving interface overtakes another
right moving interface, they give rise to a left moving interface,
$R+R\to L$, and similarly, $L+L\to R$.  The corresponding rate
equations are
\begin{eqnarray}    
   \dot R&=&-2R^2-2RL+L^2, \nonumber\\ 
   \dot L&=&-2L^2-2RL+R^2. 
\label{RL}
\end{eqnarray}
The interface concentration is readily found, 
\begin{equation}    
R(t)=L(t)={1\over 3+3t}.
\label{RLsol}
\end{equation}
The behavior is similar to the case $N=2$, as the resulting spatial
pattern form a mosaic of single-species domains whose average size is
${\cal L}\sim t$.  The previous analysis implicitly assumes that
interfaces hop one at a time, namely sequential dynamics. Alternatively,
one can consider simultaneous hoping, or parallel dynamics.  Here
interfaces move ballistically, and thus, interfaces moving with the same
velocity do not interact.  The reaction scheme is $R+L\to \emptyset$,
and the rate equations read $\dot R=\dot L=-2RL$. The resulting
interface concentrations $R(t)=L(t)=1/(3+2t)$ differ only slightly from
Eq.~(\ref{RLsol}).

\subsection{4-species}

In the 4-species model there appear static
interfaces denoted by $S$ ($AC$, $BD$, $CA$, and $DB$), in addition to
the previously defined right moving interfaces ($AB$, $BC$, $CD$, and
$DA$), and left moving interfaces ($BA$, $CB$, $DC$, and $AD$). 
Interfaces react upon collision according to the rules
$R+L\to \emptyset$, $R+S\to L$, $R+R\to S$, $L+L\to S$, and $S+L\to
R$, resulting in the following rate equations
\begin{eqnarray}    
\label{paral}   
\dot R=&-&2R^2-2RL-RS+SL, \nonumber\\
\dot L=&-&2L^2-2RL-SL+RS, \\
\dot S=&R&^2+L^2-RS-SL.   \nonumber
\end{eqnarray}
Solving these equations subject to the appropriate initial conditions
gives
\begin{equation}    
    R(t)=L(t)={1\over 4+4t}, \quad
    S(t)={1\over \sqrt{4+4t}}-{1\over 4+4t}.
\label{RLSsol}
\end{equation}
Different rules govern the decay of static and mobile interfaces, and
consequently, the coarsening process is characterized by two intrinsic
length scales.  The typical distance between two static interfaces,
$t^{1/2}$, grows slower than the distance between two moving interfaces,
$t$. A nontrivial spatial organization occurs in which large
``superdomains'' contain many domains of alternating noninteracting
($AC$ or $BD$) species, $BAAACCAAACCCAAAD$. We denote the typical
domain size by $\ell\sim t^{1/2}$ and the typical superdomain size by
${\cal L}\sim t$.  The typical number of noninteracting domains inside a
superdomain grows as ${\cal L}/\ell \sim t^{1/2}$. Such an
organization is a consequence of the existence of noninteracting
species, which first occurs at $N=4$.

It is useful to consider parallel dynamics as well. Again, the reaction
scheme is altered only in that the interfaces moving in the same
direction do not interact.  The reaction scheme, $R+L\to \emptyset$,
$R+S\to L$, and $S+L\to R$, is described by the following rate equations

\begin{eqnarray}    
\label{par}
\dot R=&-&2RL-RS+SL, \nonumber\\
\dot L=&-&2RL-SL+RS, \\
\dot S=&-&RS-SL.     \nonumber
\end{eqnarray}
Solving the rate equations we arrive at
\begin{equation}    
R(t)=L(t)=S(t)={1\over 4+2t}.
\label{solparal}
\end{equation}
Interestingly, when $N=4$ coarsening is sensitive to the details of the
dynamics.  Parallel dynamics is governed by a single length scale, in
contrast with the two scales underlying sequential dynamics.

\subsection{5-species}

In the 5-species case there are two types of stationary interfaces,
$S_R$ ($AC,BD,CE,DA,EB$) and $S_L$ ($AD,BE,CA,DB,EC$), in addition to
the right and left moving interfaces, $R$ ($AB,BC,CD,DE,EA$) and $L$
($BA,CB,DC,AD,AE$).  The reaction process is symbolized by $R+L\to
\emptyset$, $R+S_L\to L$, $R+S_R\to S_L$, $S_R+L\to R$, $S_L+L\to
S_R$, $R+R\to S_R$, and $L+L\to S_L$.  In other words, when a moving
interface hits a stationary interface of the same kind, the outcome is
a stationary interface of the opposite kind; when a moving interface
hits a stationary interface of the opposite kind, a dissimilar moving
interface emerges.  Collisions between similar moving interfaces
produce stationary interfaces of the same kind, and thus, the obvious
notations $S_L$ and $S_R$.  For the 5-species model with sequential
dynamics, the rate equations read
\begin{eqnarray}
\dot L=&-&2RL-LS_L-LS_R+RS_L-2L^2,\nonumber\\
\dot R=&-&2RL-RS_L-RS_R+LS_R-2R^2,\nonumber\\
\dot S_L=&-&RS_L-LS_L+RS_R+L^2,\nonumber\\
\dot S_R=&-&RS_R-LS_R+LS_L+R^2.
\end{eqnarray}
The reaction scheme and consequently the rate equations are invariant
under the duality transformation $(R,S_R)\longleftrightarrow (L,S_L)$.
Particularly, for $R(0)=L(0)$ and $S_R(0)=S_L(0)$, the corresponding
densities remain equal forever. This condition is certainly satisfied
for the symmetric initial conditions $R(0)=L(0)=S_L(0)=S_R(0)=1/5$.
Therefore, $R(t)= L(t)$ and $S_L(t)= S_R(t)$, and in what
follows we shall use the notations $M(=R=L)$ for mobile interfaces and
$S(=S_L=S_R)$ for stationary interfaces.  Thus, the four rate
equations reduce to a pair of rate equations:
\begin{equation}
\label{sm1}
\dot M=-4M^2-SM,\quad
\dot S=M^2-SM.
\end{equation}
These equations can be linearized by introducing a modified time
variable, $T(t)=\int_0^t M(t')dt'$. Using the notation
$'\equiv d/dT$, we rewrite the governing equations as 
\begin{equation}
M'=-4M-S,\quad 
S'=M-S.
\end{equation}
Solving these equations gives 
\begin{eqnarray}
M(T)&={1\over 5}\left(\lambda_+ e^{-\sqrt{5}\lambda_+ T}
-\lambda_- e^{-\sqrt{5}\lambda_- T}\right),\nonumber\\
S(T)&={1\over 5}\left(\lambda_+ e^{-\sqrt{5}\lambda_- T}
-\lambda_- e^{-\sqrt{5}\lambda_+ T}\right),
\end{eqnarray}
with the shorthand notations, $\lambda_{\pm}=(\sqrt{5}\pm1)/2$.  Here,
the moving interfaces are depleted at $T_{\infty}=2
(\ln\lambda_+)/\sqrt{5}$.  The density of static interfaces approaches
a finite value, $S(\infty)={1\over 5}\left(\lambda_+^{2-\sqrt{5}}
-\lambda_-^{2+\sqrt{5}}\right)\cong 0.152477$, so the average domain
size in the frustrated state is ${\cal L}(\infty)=1/2S(\infty)\cong
3.27918$.  In terms of the actual time $t$, the density of moving
interfaces decays exponentially, $R(t)\propto
e^{-S(\infty)t}$. Contrary to the previous cases, $N<5$, no coarsening
occurs and the system quickly approaches a frozen state 
of short noninteracting same-species domains separated by 
stationary interfaces. 

A similar picture is found for parallel dynamics as well.  Here, the
reaction process is $R+L\to \emptyset$, $R+S_L\to L$, $R+S_R\to S_L$,
$S_R+L\to R$, $S_L+L\to S_R$, and the rate equations are
\begin{eqnarray}
\dot L=&-&2RL-LS_L-LS_R+RS_L,\nonumber\\
\dot R=&-&2RL-RS_L-RS_R+LS_R,\nonumber\\
\dot S_L=&-&RS_L-LS_L+RS_R,\nonumber\\
\dot S_R=&-&RS_R-LS_R+LS_L.
\end{eqnarray}
The useful duality relation, $(R,S_R)\longleftrightarrow (L,S_L)$,
still applies, so there are only two independent interface
concentrations, $M$ and $S$, which evolve according to the following 
rate equations 
\begin{equation}
\dot M=-2M^2-MS,\quad
\dot S=-MS.
\end{equation}
The calculation is very similar to the sequential case, and we merely quote 
the results 
\begin{equation}
M(T)={2e^{-2T}-e^{-T}\over 5}, \quad
S(T)={e^{-T}\over 5}.
\end{equation}
The limit $t\to\infty$ corresponds to $T\to T_\infty=\ln 2$.  We find
that the density of static interfaces saturates at a finite value,
$S(\infty)=1/10$, while the density of moving interfaces decays
exponentially in time, $M(t)\propto e^{-S(\infty)t}=e^{-t/10}$.  The
average size of a domain in the frozen state is ${\cal
L}(\infty)=1/2S(\infty)=5$.

To summarize, both for parallel and sequential dynamics the rate
equations predict coarsening when the number of species is sufficiently
small, $N<5$, and fixation for a large number of species, $N\geq 5$.
When fixation occurs, each site attains a final state while for $N\leq
4$ the state of any site continues to change, although the frequency of
changes decreases with time.  It is remarkable that the rate equation
approach which neglects spatial correlations between interfaces
correctly predicts the marginal food chain length for fixation, $N_c=5$,
as has been proved rigorously for both sequential dynamics
\cite{bramson} and parallel dynamics \cite{fisch}.
 
In the coarsening cases, $N<5$, both for the two- and three-species
case the average domain size $\ell(t)$ grows linearly with time
independent of the dynamics.  In the 4-species case, however, the rate
equation theory predicts linear growth $\ell(t)\sim t$ for parallel
dynamics, and slower "diffusive" growth $\ell(t)\sim \sqrt{t}$ for
sequential dynamics. In the latter case, the larger linear scale still
exists and it characterizes the typical distance between two mobile
interfaces.

\section{COARSENING DYNAMICS IN ONE DIMENSION}

The above rate equation theory successfully predicts the fixation
transition at $N_c=5$ in agreement with rigorously known results
\cite{bramson,fisch}.  If one assumes that the average concentration of
each species is conserved throughout the process, which is clearly
correct at least in the case of equal initial concentrations, one can
find a simple argument for a lower bound on the marginal food chain
length $N_c$.  A frozen chain consists of alternating domains of
noninteracting species.  For $N=2, 3$ such a chain is impossible since
all species interact.  For $N=4$, frozen chains are filled by either $A$
and $C$ species or $B$ and $D$ species thereby violating the
conservation of the densities.  [Note, however, that for $N=4$ in {\it
finite} systems, density fluctuations could drive the system towards a
final frozen configuration].  For $N\geq 5$, a frozen chain conserving
the densities is possible and thus $N_c\geq 5$.  Given the kinetics
predicted by the mean-field rate equation approach usually proceeds with
a {\it faster} rate than actual kinetics\cite{redner}, one can
anticipate that the threshold number of different species predicted by
the mean-field theory provides an upper bound for the actual $N_c$,
$N_c\leq 5$.  This is combined with the lower bound, $N_c\geq 5$, to
yield $N_c=5$.

For $N<5$, coarsening occurs and it is quite possible that the system
develops significant spatial correlations.  In such a case, the
quantitative predictions of the rate equation theory are inaccurate.

The cyclic $N$-species Lotka-Volterra model is implemented in the 
following way. We consider a one-dimensional lattice of size ${\cal
N}$ with periodic boundary conditions. Each site $i$ of the lattice is
in a given state $N_i$ with $N_i=A,B,C,\dots$ In sequential dynamics,
we choose randomly a site and then one of its two nearest
neighbors. If the neighbor is a predator of the chosen site, the state
of the latter changes to the state of the predator.  Otherwise, the
state of the site remains the same.  Time is incremented by $1/{\cal
N}$ after each step.  For parallel dynamics, all sites are updated
simultaneously and change their state if one of their nearest
neighbors is their predator.  This cellular automata rule has been 
used in \cite{fisch} and it should be noted that the dynamics is fully
deterministic. Coarsening behavior of the system depends on spatial
fluctuations present in the initial state.  For both types of
dynamics, efficient algorithms keeping trace only of moving interfaces
have been implemented. 

Below, we present numerical findings accompanied by heuristic
arguments for the coarsening dynamics in one dimension.  Again, we
restrict ourselves to the {\it symmetric} initial
concentration.
In this case the average concentration  of each species remains $1/N$, 
despite the nonconserving microscopic evolution rules.

\subsection{2-species}

As mentioned previously, for $N=2$, interfaces perform a random walk
and annihilate upon collision.  This exactly soluble voter
model\cite{lig} is equivalent to the one-dimensional Glauber-Ising
model at zero temperature \cite{glauber,racz}. The interface
concentration is given by\cite{glauber}
\begin{equation}
I(t)=e^{-4t}[I_0(4t)+I_1(4t)]/2, 
\label{it}
\end{equation}
with $I_n(x)$ the $n^{\rm th}$ order modified Bessel function.  In the
limit $t\to 0$ correlations are absent and the interface density
$I(t)\cong 1/2-t$ agrees with the short time limit of Eq.~(\ref{i}). 
Asymptotically, the
coarsening is much slower in comparison with the rate equation 
predictions, $I(t)\simeq (8\pi t)^{-1/2}$. The system separates into
single species domains as follows
\begin{equation}
AAABBBB\underbrace{AAAA}_{\ell}BBAAA. 
\label{domains2}
\end{equation}
The average domain size $\ell$ exhibits a diffusive growth law
$\ell(t)\sim t^{\alpha}$ with $\alpha=1/2$. 

In the 2-species case, parallel dynamics is generally meaningless
since it quickly leads to a chain consisting of alternate clusters
with no more than two consecutive similar-species sites.  All sites of
the chain then change their states at each time step.  However, a
parallel realization of the sequential dynamics is possible if all
initial distances between interfaces are even integers.  In this case,
the number of interfaces gradually decreases, and a behavior very
similar to the sequential case emerges\cite{privman}.

\subsection{3-species}

It is convenient to consider first the simpler parallel
dynamics where interfaces move ballistically with velocity $\pm1$,
and annihilate upon collisions.  In this well understood ballistic
annihilation process\cite{fisch,Elskens,Krug,krl,Piasecki,Droz}, a 
simple combinatorial calculation (see subsection V.C)
yields the following interface density 
\begin{eqnarray}    
R(t)&=&{1\over 3^{2t+1}}\left[\sum_{i=0}^{t}
\left(\begin{array}{c}2t\\2i\end{array}\right)
\left(\begin{array}{c}2t-2i\\t-i\end{array}\right)\right.\nonumber\\
&&\left.\qquad +\sum_{i=0}^{t-1}
\left(\begin{array}{c}2t\\2i+1\end{array}\right)
\left(\begin{array}{c}2t-2i-1\\t-i\end{array}\right)\right].   
\label{elskens}
\end{eqnarray}
In the long time limit, the interfaces concentration decay $R\simeq
(6\pi t)^{-1/2}$ is much slower than the $t^{-1}$ decay suggested by
the rate equation (\ref{RLsol}).  The decay law governing the
interface density can be simply understood.  Consider a finite
interval of size $L$ containing interfaces with initial concentration
$c_0$. The total number of interfaces is $N=c_0L$. If the initial
conditions are random, the difference between the number of left and
right moving interfaces is roughly $\Delta N=|N_R-N_L|\sim \sqrt{N}$.
At long times, all minority interfaces are eliminated and thus, the
interface concentration approaches $\Delta N/L\sim (c_0/L)^{1/2}$.  By
identifying the box size with the appropriate ballistic length $L\sim
Vt$, the time dependent interface concentration for an infinite system
is found, $R(t)\sim (c_0/Vt)^{1/2}$.

The system organizes into large ballistically growing superdomains.
Each superdomain contains interfaces moving in the same direction,
neighboring superdomains contain interfaces moving in the opposite
direction, {\it etc}.  In addition to the average size of superdomains,
there is an additional length scale in the problem corresponding to
the distance between two adjacent similar velocity interfaces.  We
define these relevant length scales using the following illustrative
configuration
 \begin{equation}
B\overbrace{AABBB\underbrace{CCCC}_{\ell}
AAABBCCC}^{\cal L}B, 
\label{domains3}
\end{equation}
The corresponding coarsening exponents, $\alpha$ and $\beta$, are
defined via $\ell\sim t^{\alpha}$ and ${\cal L}\sim t^{\beta}$,
respectively.  For the $N=3$ with parallel dynamics we thus find
$\alpha=1/2$ and $\beta=1$.  Starting from an initially homogeneous
state, the system develops a unique spatially organized state which is
a mosaic of mosaics. Indefinitely growing superdomains contain a
growing number ${\cal D}={\cal L}/\ell\sim t^{1/2}$ of cyclically
arranged domains ($ABCABC$ and $CBACBA$).

We now turn to the complementary case of sequential dynamics.
Interfaces perform a biased random walk and thus, the ballistic motion
is now supplemented by superimposed diffusion.  In addition, two
parallel moving interfaces can annihilate and give birth to an
opposite moving interface.  It proves useful to consider the continuum
version of the model where interfaces move with velocity $+V$ and $-V$
with equal probabilities, and have a diffusivity $D$.  To establish
the long-time behavior we {\it assume} that the system organizes into
domains of right and left moving interfaces.  Inside a domain,
interfaces moving in the same direction can now annihilate via a
diffusive mechanism, unlike the parallel case.  On slower than
ballistic scales, the problem reduces to diffusive annihilation,
$X+X\to \emptyset$, where $X$ is either $R$ or $L$, with a density
decaying as $c_{\rm diff}(t)\sim (Dt)^{-1/2}$.  On ballistic scales
the problem (almost) reduces to the ballistic annihilation process
$R+L\to \emptyset$, with the density decay $c_{\rm ball}(t) \sim
(c_0/Vt)^{1/2}$ as described previously.  However, to describe the
complete ballistic-diffusion annihilation, one cannot use the initial 
concentration $c_0$ since it is constantly reduced by diffusive
annihilation.  Therefore, we replace the initial concentration $c_0$
with the time dependent concentration $c_{\rm diff}(t)$, and we find
\cite{brl,brk} $c\sim (DV^2t^3)^{-1/4}$ which in particular implies
\begin{equation}    
\ell\sim t^{3/4}. 
\label{eli}
\end{equation}
This result is quite striking since separately both annihilation
processes, diffusion-controlled and ballistic-controlled, give the
same coarsening exponent 1/2, so one expects that their combination does
not change the behavior while in fact it enhances the coarsening exponent
to 3/4. The resulting spatial structure is similar to the parallel
case, Eq.~(\ref{domains3}). However, the smaller length scale is now a
geometric average of a diffusive and a ballistic scale as  follows
from Eq.~(\ref{eli}), while the larger scale remains unchanged, 
${\cal L}\sim t$.

We have performed Monte Carlo simulations for 100 realizations on a 
lattice of size $10^6$, for times up to $t\simeq 10^6$. Results are
shown in the Fig.~1. The interface concentration decays algebraically,
$R\propto t^{-\alpha}$ with an exponent $\alpha\cong 0.79$.  A careful
analysis shows that the local slope $\alpha(t)=d \ln c(t)/d\ln t$
should asymptotically approach the value $-3/4$. It is possible that
this finite time effect can be attributed to the recombination
reaction ($R+R\to L$), which is not an annihilation
reaction. Nevertheless, a single $L$-interface inside an $R$-domain is
quickly annihilated by nearest $R$-interface, and therefore
recombination is asymptotically equivalent to annihilation. We expect
that as $t\to \infty$, the coarsening exponent is indeed $\alpha=3/4$.

\begin{figure}
\narrowtext
\epsfxsize=\hsize
\epsfbox{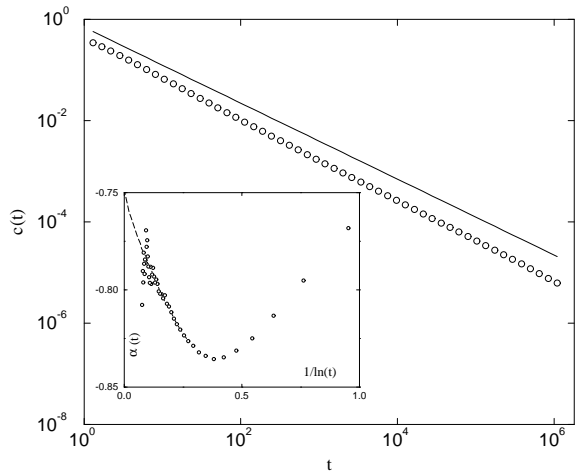}
\caption{The concentrations of interfaces as a function of time
for the 3-species model and a sequential dynamics in a log-log plot.
A line of slope $3/4$ is shown as a reference. The insert shows the local
exponent $\alpha(t)$ as a function of $1/\ln t$. A limiting
value of $\alpha\rightarrow 3/4$ is plausible.
\label{fig1}}
\end{figure}
 
To summarize, the resulting spatial patterns in the $N=3$ case consist
of superdomains of cyclically arranged domains as in Eq.~(\ref{domains3}).
The larger length scale is ballistic, ${\cal L}\sim t$, while the
smaller length scale is sensitive to the microscopic details of the
dynamics: $\ell\sim t^{1/2}$ for parallel dynamics and $\ell\sim
t^{3/4}$ for sequential dynamics.

\subsection{4-species}

For the 4-species cyclic Lotka-Volterra model, numerical simulations
indicate that parallel and sequential dynamics are asymptotically
equivalent and that the domain structure is qualitatively similar to
the predictions of the sequential rate equations, $M(t)\ll S(t)$.  We
use heuristic arguments to obtain the values of the coarsening
exponents $\alpha$ and $\beta$, characterizing the density decay of
mobile $M(t)\sim t^{-\alpha}$, and static $S(t)\sim t^{-\beta}$
interfaces. 

Given the equivalence of parallel and sequential dynamics, we restrict
ourselves to the simpler former dynamics.  What is the spatial
structure in the long time limit?  Since $M(t)\ll S(t)$ (here and
below $M(t)=R(t)=L(t)$ denotes the density of moving interfaces), we 
assume an alternating spatial structure of ``empty'' regions (with no
more than one moving interface) and ``stationary'' regions (with many
stationary interfaces inside any such region).  If the 
interface densities obey scaling, then the size of the empty and the
stationary regions should be comparable.  The typical size of an empty
or a stationary region is therefore of the order of $M^{-1}$.  The
typical number of stationary interfaces inside a stationary region is
of the order of $S/M$.  The evolution proceeds as follows: A moving
interface hits the least stationary particle and bounces back (since
$R+S\to L$ and $S+L\to R$). Then this interface hits the least
stationary particle of the neighboring stationary region, and bounces
back again.  This ``zig-zag'' process continues and at some time one
of these stationary regions ``melts'', thereby giving birth to a
larger empty region.  If there is a moving particle inside merging
empty region, the two moving particles quickly annihilate.  If there
is no such particle, the moving particle continues to eliminate
stationary interfaces. This process is illustrated in Fig.~2.  

\begin{figure}
\narrowtext
\epsfxsize=\hsize
\epsfbox{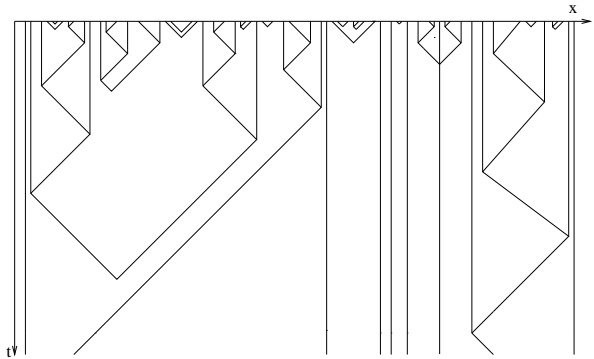}
\caption{Space time diagram of the interface motion in the 
4-species case with parallel dynamics. 
\label{fig2}}
\end{figure}

The
typical time $\tau$ for a stationary region to melt is
$\tau=M^{-1}\times S/M=S/M^2$.  This melting time $\tau$ is also the
typical time for annihilation of a moving interface and thus,
\begin{equation}    
\dot M\sim -{M\over \tau}\sim -{M^3\over S}.
\label{M}
\end{equation}
Substituting $S(t)\propto t^{-\alpha}$ and $M(t)\propto t^{-\beta}$
into Eq.~(\ref{M}) we get the exponent relation 
\begin{equation}    
2\beta-\alpha=1. 
\label{laur}
\end{equation}

In the next section, we introduce the mutation distribution, and find 
an equivalence between the fraction of persistent sites and the static 
interface density. Using this relation and a simple solvable example, 
we find the exponent relation 
\begin{equation}    
\alpha+\beta=1. 
\label{relation}
\end{equation}
The two exponent relations therefore imply the values
$\alpha=1/3$ and $\beta=2/3$. We have simulated 100 systems of size
$10^6$ up to times $t\simeq 10^6$. The results are shown in Figs.~3
and 4.  We have found $S(t)\propto t^{-0.34}$, $R(t)=L(t)\propto
t^{-0.69}$ for parallel dynamics, and $S(t)\propto t^{-0.35}$,
$R(t)=L(t)\propto t^{-0.70}$ for sequential dynamics.  We conclude
that the simulation results support the above predictions.

As in the
three-species case there are two relevant growing length scales. The
system organizes into domains of alternating noninteracting species
with a typical size $\ell\sim t^{1/3}$. On the other hand, active
interfaces are separated by a typical distance of ${\cal L}\sim
t^{2/3}$, according to the following illustration 
\begin{equation}
B\overbrace{AACCCAAA\underbrace{CCCC}_{\ell}
AACCAAACCC}^{{\cal L}}D. 
\label{domains4}
\end{equation}
In any finite lattice, density fluctuations drive the system towards a
final frozen or ``poisoned'' configuration, {\it i.e.}  configuration
filled by either $A$ and $C$, or $B$ and $D$.  This poisoning happens
when the size of superdomains becomes of the order of lattice size.
The poisoning time is therefore proportional to ${\cal N}^{3/2}$ for
an ${\cal N}$-site chain.
\begin{figure}
\vspace{-.1in}
\narrowtext
\epsfxsize=\hsize
\epsfbox{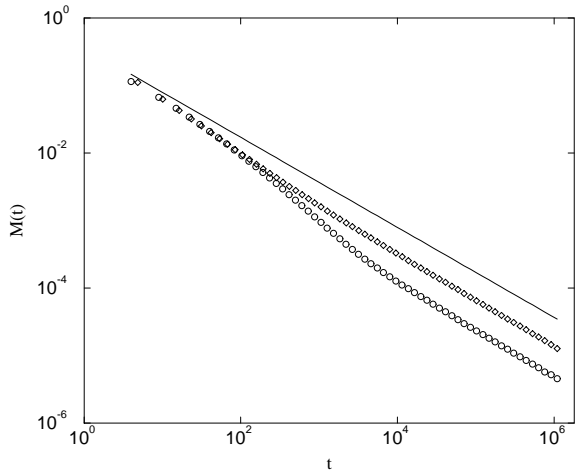}
\vspace{-.1in}
\caption{The concentrations of the moving interfaces as a
function of time for the 4-species cyclic Lotka-Volterra model with
sequential dynamics (diamonds) and with parallel dynamics
(circles). The slope give the exponents $\beta_{\rm seq}=0.70$ and
$\beta_{\rm par}=0.69$.  A line of slope $2/3$ is shown as a
reference.
\label{fig3}}
\end{figure}
\begin{figure}
\vspace{-.1in}
\narrowtext
\epsfxsize=\hsize
\epsfbox{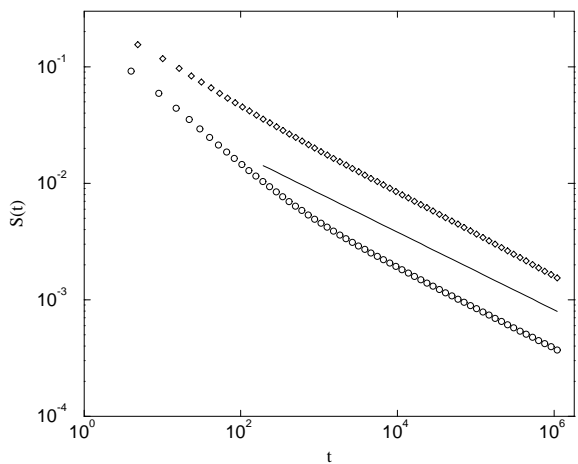}
\vspace{-.1in}
\caption{
The concentrations of the stationary interfaces as a
function of time for the 4-species model for sequential dynamics
(diamonds) and parallel dynamics (circles). The slope give the
exponents $\alpha_{\rm seq}=0.35$ and $\alpha_{\rm par}=0.34$.
A line of slope $1/3$ is shown as a reference.  
\label{fig4}}
\end{figure}

We stress that a 3-velocity ballistic annihilation model, $R+L\to
\emptyset, R+S\to \emptyset, S+L\to \emptyset$, has been recently
investigated \cite{krl,Piasecki,Droz,brl}, and the symmetric case,
$R\equiv L$, has been solved exactly \cite{Droz}.  For the special
initial condition, $R(0)=L(0)=3S(0)/2$, a surprisingly similar
behavior $R(t)\sim S(t)\sim t^{-2/3}$ occurs.  It would be interesting
to establish a relationship between this solvable ballistic
annihilation model and the interface motion in the 4-species process.

\subsection{5-species}

For the 5-species cyclic Lotka-Volterra model, it is well known that
the system approaches a frozen state\cite{bramson,fisch}.  The
kinetics approach towards saturation has not been established, though.

We now present a heuristic argument for estimating the concentration
decay of the mobile interfaces.  Since the density of mobile
interfaces rapidly decreases while the density of stationary
interfaces remains finite we can ignore collisions between mobile
interfaces.  Thus we should estimate the survival probability of a
mobile interface in a sea of stationary ones.  There are two reactions
in which moving interfaces survive although they change their type,
$R+S_L\to L$ and $L+S_R\to R$.  Thus, a right moving interface is long
lived in the following environment
\begin{equation}
\cdots S_RS_RS_RS_RMS_LS_LS_LS_L\cdots
\label{domains5}
\end{equation}
Clearly, in such configurations the zig-zag reaction process takes
place.  The moving interface travels to the right during a time
$t_0\sim 1/c_0v_0$, eliminates a stationary interface and travels to
the left a time of order $2t_0$, eliminates an interface and travels
back to the right, {\it etc}. Thus, to eliminate $N_s$ interfaces, the
moving interface should spend a time of order $t\simeq
t_0\sum_{i=1}^{N_s}i=t_0N_s(N_s+1)/2$.  Therefore, the number of
stationary interfaces $N_s(t)$ eliminated by a moving interface scales
with time as $N_s(t)\sim \sqrt{c_0v_0t}$.  Configurations of the type
(\ref{domains5}) are encountered with probability $\propto e^{-N_s}$
with $N_s$ the configuration length, and thus, the density of moving
interfaces exhibits a stretched exponential decay,
\begin{equation}
M(t)\propto e^{-{\rm const.}\times \sqrt{c_0v_0t}}.
\label{stretched}
\end{equation}

The stretched exponential behavior (\ref{stretched}) is expected 
to appear for arbitrary
$N\geq 5$.  When the number of interfaces exceeds the threshold value,
$N>5$, stationary interfaces of ``intermediate'' types arise, the
crossover from initial exponential behavior to the asymptotic
stretched exponential behavior is shifted to larger times and
therefore harder to observe numerically.  For the threshold number of
species, however, we have found a convincing agreement between the
theoretical prediction of Eq.~(\ref{stretched}) and numerical results
(see Fig.~5).  Finally we note that the actual kinetics
(\ref{stretched}) is {\it slower} than the mean-field counterpart,
$M_{\rm MFT}(t)\propto e^{-t}$, due to spatial correlations.

\begin{figure}
\narrowtext
\epsfxsize=\hsize
\epsfbox{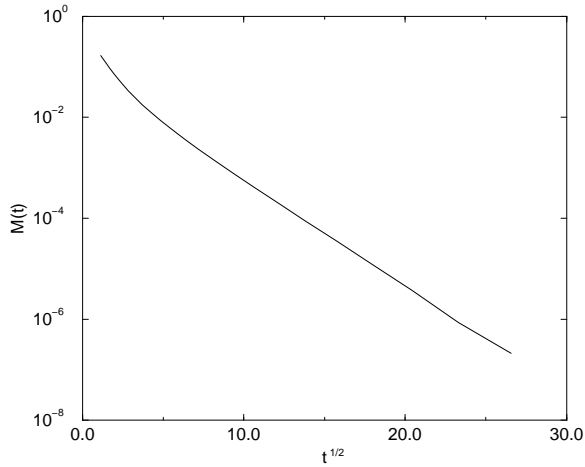}
\caption{The concentration of moving interfaces as a
function of the square root of time in linear-log plot 
for the 5-species cyclic Lotka-Volterra model with
sequential dynamics.
\label{fig5}}
\end{figure}

\section{DYNAMICS OF MUTATIONS}

Consider a lattice site occupied by some species, say $A$.  What is
the probability that this site has been occupied by the same species
during the time interval $(0,t)$?  Otherwise, what is the fraction of
$A$ sites which never ``mutated''?  We denote the fraction of
``persistent'' $A$ species by $A_0(t)$; $B_0(t)$, $C_0(t)$, {\it etc.}
are defined analogously.  We can further generalize these
probabilities to define, {\it e.g.}, $A_n(t)$, the fraction of sites
that have undergone exactly $n$ mutations during the time interval
$(0,t)$.  We start by analyzing these quantities on the mean-field
level and then describe exact, scaling, and numerical results in one
dimension.

\subsection{Mean-Field Theory}

Let us investigate the 3-species cyclic Lotka-Volterra on the complete
graph; the generalization to the $N$-species case is straightforward.
The rate equations describing the evolution of the mutation
distribution $A_n(t)$ are 
\begin{eqnarray}    
\label{muta}
\dot A_{3n}=&  aA_{3n-1}-cA_{3n},\nonumber\\
\dot A_{3n+1}=&cA_{3n}-bA_{3n+1},\\
\dot A_{3n+2}=&bA_{3n+1}-aA_{3n+2} \nonumber
\end{eqnarray}
with $A_{-1}(t)\equiv 0$.
Analogous equations can be written for $B_n(t)$ and $C_n(t)$ by 
cyclic permutations.  These rate equations form an infinite set of
linear equations with $a(t), b(t)$, and $c(t)$ as (time-dependent)
coefficients.  Therefore, the general case is hardly tractable
analytically since the coefficients, {\it i.e.}  solutions of
Eqs.~(\ref{abc}), are elliptic functions.  We therefore restrict our
attention to the symmetric case $a=b=c=1/3$, and examine $P_n(t)$, the
total fraction of sites mutated exactly $n$ times.  The quantity $P_n(t)$
evolves according to
\begin{equation}
\label{mmfe}
\dot P_n=P_{n-1}-P_n, 
\end{equation}
with $P_{-1}\equiv 0$ to ensure $\dot P_0=-P_0$.  In Eq.~(\ref{mmfe})
we absorbed the concentration factor 1/3 into the time-scale for
convenience.  Solving (\ref{mmfe}) subject to the initial condition
$P_n(0)=\delta_{n0}$, one finds a Poissonian mutation distribution
\begin{equation}
\label{mmfs}
P_n(t)={t^n\over n!}\,e^{-t}.
\end{equation}
This mutation distribution is identical to the one found for the voter
model\cite{elp} on the mean-field level.

The distribution is peaked around the average $\langle n \rangle=t$,
and the width of the distribution, $\sigma$, is given by 
$\sigma^2=\langle n^2 \rangle-\langle n\rangle=t$.  In the limits,
$t\to \infty$, $n\to \infty$, and $(n-t)/\sqrt{t}$ finite, $P_n(t)$
approaches a scaling form
\begin{equation}
\label{mmfscaling}
P_n(t)={1\over\sigma}
\Phi_{\infty}\left({n-\langle n\rangle\over\sigma}\right), 
\end{equation}
where the scaling distribution function $\Phi_{\infty}(z)$ is
Gaussian, $\Phi_{\infty}(z)=(2\pi)^{-1/2} \exp(-z^2/2)$, and the index
$\infty$ indicates that the solution on the complete graph correspond
to the infinite-dimensional limit.  We also note that the fraction of
persistent sites decreases exponentially, $P_0(t)=e^{-t}$.

Let the initial state of a site be $A$, without loss of generality,
then the probability that the state is $A$ at time $t$ is given by
$R_0(t)=\sum_n P_{3n}(t)$. In general, three such autocorrelation
functions
\begin{equation}
R_k(t)=\sum_{n=0}^{\infty}P_{3n+k}(t)
\end{equation}
correspond to the three possible outcomes at time $t$, $A$ if $k=0$,
$C$ if $k=1$, and $B$ if $k=2$. The quantity $R_0(t)$ is evaluated from
equation (\ref{mmfs}) using the identity $e^t+e^{\zeta t}+e^{\zeta
^2t}=3\sum_n t^{3n}/(3n)!$, with $\zeta=e^{2\pi i/3}$. Generally, we find
that
\begin{equation}
R_k(t)={1\over3}\left[1+2e^{-3t/2}\cos\left({\sqrt{3}\over 2}t
+{4\pi k\over 3}\right)\right],
\end{equation}
for $k=0,1,2$.  The structure of the autocorrelation functions is
rather simple -- an exponential approach to the equilibrium value
$R_k(\infty)=1/3$ is accompanied by oscillations. 
The three autocorrelation functions
differ only by a constant phase shift.  One can verify that
exponential decay occurs for arbitrary $N$, and that temporal
modulations occur when $N>2$. Numerical simulations for the case $N=3$
indeed show temporal oscillations in $R_0(t)$. However, algebraic
rather than exponential decay is found for the magnitude.

\subsection{Scaling behavior}

Mutation dynamics and coarsening dynamics are closely
related\cite{elp}.  For example, the rate of mutation is given by the
density of moving interfaces.  Using similar scaling arguments, we
study asymptotic properties of the mutation distribution in the
one-dimensional case.

The mutation distribution satisfies the normalization condition,
$\sum_n P_n=1$. Let the average number of mutations be $\langle n
\rangle=\sum_n n P_n$. Every motion of an interface contributes to an
increase in the number of mutations in one site, and thus the mutation
rate equals the density of moving interfaces, $d\langle
n(t)\rangle/dt=M(t)$. In the coarsening case, $N<5$, we found that the
moving interface density decays algebraically, $M(t)\sim
t^{-\mu}$. Therefore, the average number of mutations grows
algebraically, $\langle n(t) \rangle \sim t^{\nu}$, with $\nu=1-\mu$.
For $N=2$ and 3, the density of moving interfaces decays
inversely proportional to the average domain size, $M\sim
\ell^{-1}$, since stationary interfaces are absent when $N\leq
3$; therefore, $\mu=\alpha$. For $N=4$, however, the density of moving
interfaces is inversely proportional to the average size of
superdomains, $M\sim {\cal L}^{-1}$, implying $\mu=\beta$.

In the case of $N=2$, it has been shown that the mutation distribution
obeys scaling \cite{elp}.  We assume that this behavior generally
holds when the system coarsens,
\begin{equation}
P_n(t)={1\over \langle n(t)\rangle}\,
\Phi\left({n\over \langle n(t)\rangle}\right). 
\label{scal}
\end{equation} 

The behavior of the scaling function, $\Phi(z)$, in the limit of small
and large arguments $z$ reflects the fraction of persistent and
rapidly mutating sites, respectively. Typically, the fraction of
persistent sites decays algebraically in time, $P_0(t)\sim
t^{-\theta}$, with $\theta$ the persistence exponent.  This exponent
has been studied recently in several contexts such as kinetic spin
systems with conservative and non-conservative dynamics and
diffusion-reaction systems \cite{dbg,kbr,der,stev,bdg,cardy,ms}.  In
the $N=2$ case, the scaling function was found to be algebraic,
$\Phi(z)\sim z^{\gamma}$, in the limit $z\to 0$.  Assuming this
algebraic behavior for $N=3$ and 4 as well implies the exponent relation
\begin{equation}
\theta=\nu(\gamma+1).
\label{tng}
\end{equation}
 
The large $z$ limit describes ultra-active sites. A convenient way to
estimate the fraction of such sites is to consider sites which make of
the order of one mutations per unit time.  At time $t$, the fraction
of these rapidly mutating sites is exponentially suppressed,
$P_t(t)\propto \exp(-t)$.  It is therefore natural to assume the
exponential form $\Phi(z)\sim \exp(-z^{\delta})$ for the tail of the
scaling distribution, thereby implying an additional exponent relation
$\mu\delta=1$. To summarize, the scaling function underlying the
mutation distribution has the following limiting behaviors
\begin{equation}
\Phi(z)\sim\cases{
z^{\gamma}&$z\ll 1$;\cr
\exp(-{\rm const.}\times z^{\delta})&$z\gg 1$.}
\label{phi}
\end{equation}

In section IV, we obtained the exponent $\mu$, characterizing the
decay of moving interfaces.  The mutation exponent and the tail
exponent are readily found using the respective exponent relations,
$\nu=1-\mu$ and $\delta=1/\mu$. To find the persistence exponent, we
note the equivalence between the fraction of persistent sites and the
fraction of unvisited sites in the interface picture \cite{kbr,cardy}.
For $N=2$, the value $\theta=3/8$ has been established analytically
\cite{der}.  For $N=3$, different behaviors were found for parallel
and sequential dynamics, and therefore it is necessary to distinguish
between the two cases. As mentioned above, the parallel case reduces
to a two-velocity ballistic annihilation process.  The probability
that a bond has remained uncrossed from the left by right-moving
interfaces is $S_+(t)\sim t^{-1/2}$, see Eq.~(\ref{elskens}).
Analogously, $S_-(t)\sim t^{-1/2}$, and consequently
$P_0(t)=S_-(t)S_+(t)\sim t^{-1}$ or $\theta=1$ follows \cite{krl}. In
the sequential case, we have not been able to determine the
persistence exponent analytically, and a preliminary numerical
simulation suggests that $\theta=1$ as in the parallel case.

For $N=4$, the number of unvisited sites is equivalent asymptotically
to the survival probability of a static interface, $P_0(t)\sim S(t)$,
and using the definitions of section IV, we find $\theta=\alpha$, {\it
i.e.} $\theta=1/3$.  We now present a heuristic argument supporting
the exponent relation (\ref{relation}).  Substituting the previously
established exponent relations $\nu=1-\mu$, $\mu=\beta$, and
$\theta=\alpha$ in Eq.~(\ref{tng}) yields
\begin{equation} 
\alpha=(1-\beta)(\gamma+1).
\label{abg}
\end{equation}

We now argue that $\gamma=0$ and thus Eq.~(\ref{abg}) reduces to
$\alpha+\beta=1$, {\it i.e.} to Eq.~(\ref{relation}). We first recall
that interfaces in
the 4-species case react according to $R+S\to L$, $L+S\to R$, and
$R+L\to \emptyset$. In the long-time limit, the zig-zag reactions
$R+S\to L$ and $L+S\to R$ dominate over the annihilation reaction
$R+L\to \emptyset$.  We, therefore, consider a simpler solvable case
where a single mobile interface is placed in a regular sea of static
interfaces.  Then this interface moves one site to the right, two to
the left, three to the right {\it etc.}  In a time interval $(0,t)$,
this interface eliminates $N_s\sim t^{1/2}$ static interfaces.  The
origin is visited $N_s$ times, site $1$ is visited $N_s-1$, site $-1$
is visited $N_s-2$, {\it etc.}  This implies that the mutation
distribution is $P_n(t)=\langle n\rangle^{-1}\Phi(n/\langle
n\rangle)$, with $\langle n\rangle\sim N_s$ and $\Phi(z)=1$ for $z<1$
and $\Phi(z)=0$ for $z>1$. Hence, ignoring the annihilation reaction
leads to $\gamma=0$. This approximation is inappropriate for
predicting the tail of $\Phi(z)$ which is sensitive to annihilation of
the moving interfaces.  However, in the small $z$ limit the 
annihilation process should be negligible, and thus $\gamma=0$.

Monte-Carlo simulations confirm the anticipated scaling behavior of
Eq.~(\ref{scal}). In Fig.~6, the scaled mutation distribution function
$\langle n\rangle P_n(t)$ is plotted versus the scaled mutation number
$n/\langle n\rangle$, for a representative case $N=4$ at different
times $t=10^3,10^4,10^5$. It is seen that the plots are time
independent. Furthermore, the scaling function approaches a finite
nonzero value in the limit of small $z=n/\langle n\rangle$, in
agreement with the scaling predictions, $\gamma=0$. 

\begin{figure}
\narrowtext
\epsfxsize=\hsize
\epsfbox{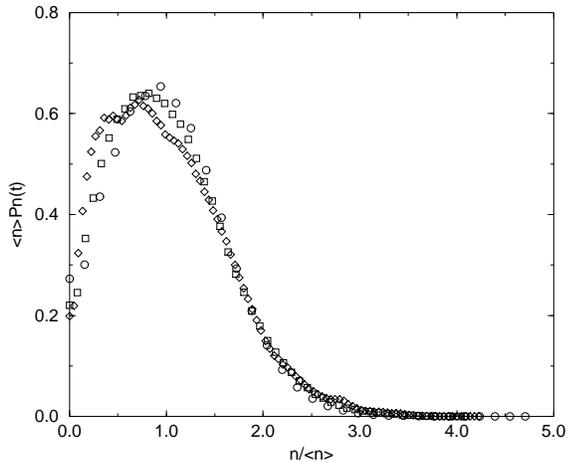}
\caption{The mutation distribution scaling function for 
a representative case of $N=4$ with sequential dynamics.
Simulations of 10 realizations of a system of size $10^6$ for time
$t=10^3$ (circles), $t=10^4$ (squares) and $t=10^5$ (diamonds).
\label{fig6}}
\end{figure}

In summary, coarsening dynamics can be characterized by a set of
exponents $\alpha, \beta, \gamma, \delta, \nu, \theta$. Table 1
gives the values of these exponents which are believed to be exact,
although for some of the exponents only numerical evidence exists so
far. 

\vspace{.2in}
\centerline{
\begin{tabular}{|l|c|c|c|c|c|c|}
\hline
N&$\alpha$&$\beta$&$\nu$&$\delta$&$\theta$&$\gamma$\\ 
\hline
2             &\,1/2\,&   &\,1/2\,&\,2\,&\,3/8\,&\,-1/4\,\\
3 (parallel)  &\,1/2\,&\,1\,&\,1/2\,&\,2\,&\,1\,&\,1\,\\
3 (sequential)&\,3/4\,&\,1 \,&\,1/4\,&\,4\,&\,1\,&\,1/3\,\\
4             &\,1/3\,&\,2/3\,&\,1/3\,&\,3\,&\,1/3\,&\,0\,\\
4 (symmetric)  &\,3/8\,&\,1/2\,&\,1/2\,&\,2\,&\,3/8\,&\,-1/4\,\\
\hline
\end{tabular}}
\vspace{.1in}
{\small Table 1 Coarsening and mutation exponents in 1D.} 
\vspace{.1in}
\subsection{An exactly solvable case}

The 3-species Lotka-Volterra model with parallel dynamics is
equivalent to the exactly solvable two-velocity ballistic annihilation
\cite{Elskens}.  We exploit this equivalence to compute analytically
the mutation distribution.  A species in a given site mutates each
time it is crossed by an interface.  As the fraction of persistent
sites is equivalent to the fraction of uncrossed bonds, the fraction
of sites visited $n$ times equals the fraction of bonds crossed
exactly $n$ times by the interfaces.  In the symmetric case, the
initial concentration of moving interfaces of velocity $+1$ or $-1$ is
1/3 (interfaces are initially absent with probability $1/3$).
Interfaces move ballistically and the system is deterministic, {\it
i.e.}, any late configuration is a one-to-one function of the initial
configuration.  It is also natural to consider integer times $t$.  The
distribution $P_n(t)$ for a given site is completely determined by the
initial distribution of the interfaces on the $t$ bonds to the left of
this site and on the $t$ bonds to the right of this site since, further
interfaces can not reach the site in a time $t$.  This $2t$ initial
bonds can be mapped onto a random walk with uncorrelated steps of
length $\pm 1$ or zero since interfaces are initially uncorrelated.
We set $S_0=0$ and define $S_i$ recursively via
$S_i=S_{i-1}+v_i,\,i=1,\ldots,t$ where $v_i=\pm 1$ is the velocity of
the $i^{\rm th}$ interface to the right of the considered site and
$v_i=0$ if the interface is absent.  Similarly, 
$S_{-i}=S_{-(i-1)}-v_{-i},\,i=1,\ldots,t$.  Thus, one has two random
walks starting from the origin, $(i,S_i)$ and $(-i,S_{-i}),
i=0,\ldots,t$, with $i$ being a time-like variable and $S_i$ the
displacement.  The crucial point is that the number of interfaces
crossing the target site at the origin during the time interval
$(0,t)$ is given by the absolute value of the minimum of the combined
random walk $(i,S_i), i=-t,\ldots,t$ (see Fig.~7). 

\begin{figure}
\narrowtext
\epsfxsize=\hsize
\epsfbox{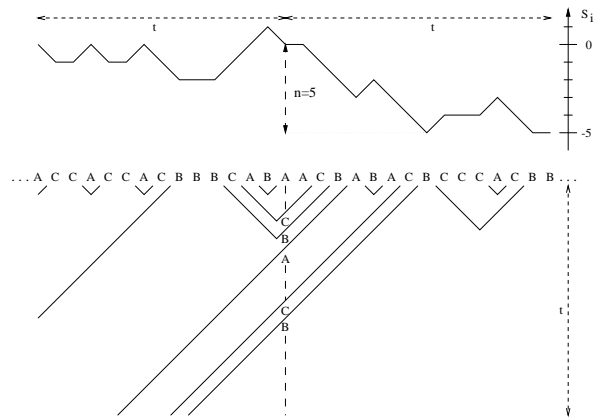}
\caption{Mapping of the initial distribution of the species on a
random walk. The absolute minimum reaches by the random walk is equal to
the number of mutations undergoes by the center site.
\label{fig7}}
\end{figure}

Indeed, the
minimum attained by the random walker on the left (right) gives the
excess of interfaces coming from the left (right) not destroyed by
other left (right) interfaces that would cross the considered site.
Thus, $P_n(t)$ equals the probability that the minimum of two
independent $t$-steps random walks starting at $S_0=0$ is $-n$,
\begin{equation}    
P_n(t)=2Q_n(t)\sum_{k=0}^n Q_k(t) -Q_n(t)^2. 
\label{ptn}
\end{equation}
The probability $Q_n(t)$ that a $t$-steps random walk starting 
at the origin has a minimum at $-n$, 
is given by\cite{feller}
\begin{equation}    
Q_n(t)=\tilde Q_n(t)+\tilde Q_{n+1}(t),
\label{ptn1}
\end{equation}
with
\begin{equation}    
\tilde Q_n(t)=
{1\over 3^t}\sum_{i=0}^{t-n}{t!\over i! \left({t+n-i\over 2}\right)!
\left({t-n-i\over 2}\right)!}. 
\label{ptn2}
\end{equation}
The trinomial coefficient in the above sum is set to zero if
$(t-n-i)/2$ is not an integer.  The sum in the right-hand side of
Eq.~(\ref{ptn}) gives the probability that the other walker has its
minimum at $-k$, with $k\leq n$, the factor 2 reflects the fact that
there are two random walkers. We subtract the last quantity $Q_n(t)^2$
which has been counted twice in the summation.  In particular, we have
$P_0(t)=Q_0(t)^2$, in agreement with the argument of the previous
section.  Also, one can show that the density of moving interfaces can
be expressed via $Q_0$ as $R(t)=Q_0(2t)/3$, leading to
Eq.~(\ref{elskens}).

To determine the asymptotic behavior of $P_n(t)$ we first compute
$\tilde Q_n(t)$.  Making use of the Gaussian approximation for the
trinomial coefficients we find
\begin{equation}    
\tilde Q_n(t)\simeq 
\sqrt{3\over 4\pi t}\,e^{-3n^2/4t}.
\label{gauss}
\end{equation}
Combining (\ref{gauss}), (\ref{ptn1}), and (\ref{ptn}) yields
\begin{equation}
P_n(t)\simeq 
\sqrt{12\over \pi t}\,{\rm Erf}\left({n\over \sqrt{4t/3}}\right)\,
e^{-3n^2/4t}, 
\label{pntt}
\end{equation}
with ${\rm Erf}(z)={2\over\sqrt{\pi}}\int_0^z du e^{-u^2}$. 
The existence of an exact solution is very useful for testing the 
validity of the scaling assumption. Indeed, Eq.~(\ref{pntt}) agrees with
the general scaling form of Eq.~(\ref{scal}), and the corresponding 
scaling function is 
\begin{equation}
\Phi(z)={4\over \sqrt{\pi}}\,e^{-z^2}{\rm Erf}(z),
\label{scalpntt}
\end{equation}
with the scaling variable $z=n/\sqrt{4t/3}$. The limiting behavior of
this scaling function agrees with the predictions of Eq.~(\ref{phi})
as well,
\begin{equation}
\Phi(z)\sim\cases{
z&$z\ll 1$;\cr
e^{-z^2}&$z\gg 1$.}
\label{phiz}
\end{equation}
The corresponding values of the scaling exponents $\nu=\mu=1/2$,
$\delta=2$, $\theta=1$, and $\gamma=1$, are in agreement with Table 1.

\section{EXTENSIONS} 

The cyclic lattice Lotka-Volterra model can be generalized in a number
of directions.  A natural generalization is to higher dimensions.
Two-dimensional case seems to be especially interesting from the point
of view of mathematical biology.  In the exactly solvable $N=2$ case
(the voter model), coarsening occurs for $d\leq 2$\cite{lig}, for the
marginal dimension $d=2$, the density of interfacial bonds decays
logarithmically, $c(t)\sim 1/\ln t$\cite{lp}, while for $d>2$, no
coarsening occurs and the system reaches a reactive steady state.  In
two dimensions, our numerical simulations indicate that there is no
coarsening, {\it i.e.} the density of reacting interfaces saturates at a
{\it finite} value.  For sufficiently large number of species
the fixation is expected but we could not determine the threshold value,
at least up to $N=10$ we have seen no evidence for fixation.

Below, we mention few other possible generalizations
and outline some of their attendant consequences.

\subsection{Asymmetric Initial Distribution}

We consider uncorrelated initial conditions with unequal species
densities.  Even in the 2-species situation, the behavior is
surprisingly non-trivial.  In particular, the densities of both
species remain constant; the persistence exponent $\theta_A$ decreases
from 1 to 0 as the initial concentration $a_0$ increases from 0 to 1
\cite{elp,der}, with $\theta_A=\theta_B=3/8$ for equal initial
concentrations\cite{der}.

Turn now to the 3-species case and consider first parallel
dynamics. In general, the densities of right and left moving
interfaces are equal as well.  However, the initial interface
distribution is correlated in the general asymmetric case and
therefore the equivalence to ballistic annihilation is less useful.
We find numerically that the interface density exhibits the same decay
as in the symmetric case, $c(t)\sim t^{-1/2}$.  To illustrate this
property let us consider the following example where the initial
densities are $a_0=1-2\epsilon$ and $b_0=c_0=\epsilon$, with
$\epsilon\to 0$.  Initially, the $A$-species dominates over the two
minority species.  While isolated $B$'s are immediately eaten by the
neighboring $A$'s, $C$-species domain arise and soon the $C$'s
dominate the system.  However, the ultimate fate of the system is
determined by pairs of nearest neighbors which are dissimilar
minorities, {\it i.e.} $BC$ and $CB$.  Initially, these interfaces are
present with probability $\epsilon^2$; clearly, they are long-lived
right and left moving interfaces. These interfaces are uncorrelated
and thus their density decays as $t^{-1/2}$.  We also performed
numerical simulations for the 3-species cyclic Lotka-Volterra model
with sequential dynamics, and the interface decay, $c(t)\sim
t^{-3/4}$, was found similar to the symmetric case.  The
interface concentration does not provide a complete picture of the
spatial distribution.  The main difference with the $3$-opinions voter
model is that the species densities are not conserved, and they
exhibit a more interesting behavior (see Fig.~8).  It is possible that
the limit where one species initially occupies a vanishingly small
volume fraction is tractable analytically, similar to recent studies
\cite{elp,stev} of Glauber and Kawasaki dynamics. 

\begin{figure}
\vspace{-.1in}
\narrowtext
\epsfxsize=\hsize
\epsfbox{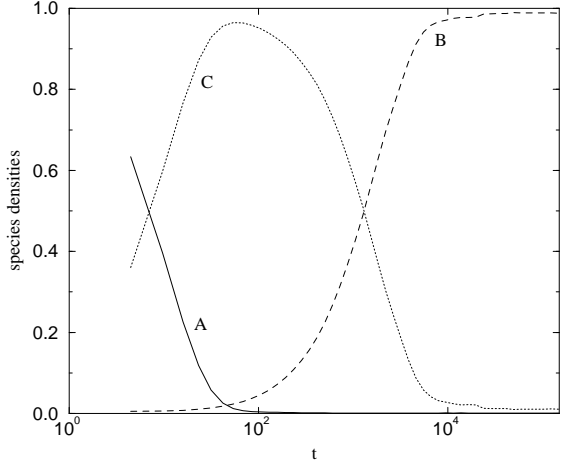}
\vspace{-.1in}
\caption{ The species densities as a function of $\ln t$ 
for the 3-species cyclic model with unequal initial densities
($a_0=0.9$ and $b_0=c_0=0.05$).
\label{fig8}}
\end{figure}
\begin{figure}
\vspace{-.1in}
\narrowtext
\epsfxsize=\hsize
\epsfbox{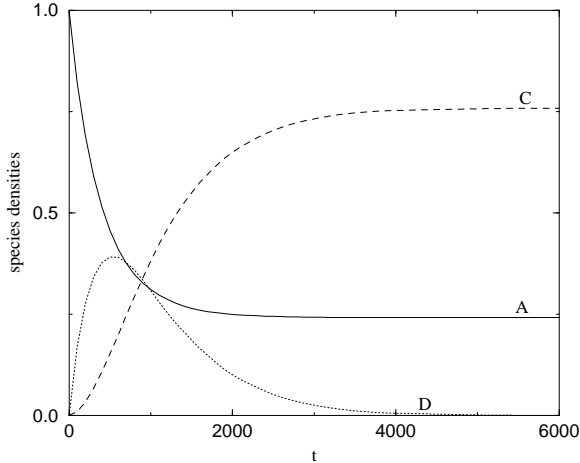}
\vspace{-.1in}
\caption{  The species densities as a function of time
for the 4-species cyclic model with unequal initial densities
($a_0=0.997$ and $b_0=c_0=d_0=0.001$).
\label{fig9}}
\end{figure}

Consider now the 4-species model. Numerically, we observed a rich
variety of different kinetic behaviors.  Rather than giving a complete
description, we restrict ourselves to a few remarks based on simulation
results and heuristic arguments.  First, the species densities are not
conserved globally, in contrast with the symmetric initial conditions or
the ordinary 4-opinions voter model.  Furthermore, if the initial
densities are different, the system can fixate and thus reach a state
such as $AAACCCACCA$ where the evolution is frozen.  In order to
illustrate the rich behavior of this system we consider the following
initial conditions $a_0=1-3\epsilon$ and $b_0=c_0=d_0=\epsilon$ with
$\epsilon\to 0$.  Eaten by the dominant $A$'s and with almost no preys,
$B$'s quickly disappear from the system. The $D$'s are growing because
they have much food and almost no predators. After a while, the $C$'s
also have some food and no predators and they overtake the $D$'s. The
$A$'s are eaten first but once the $C$'s dominate the $D$'s, $A$'s have
less and less predators.  The concentration of $D$-species and the
density of the moving interfaces decay exponentially and, therefore, the
system quickly reaches a frozen state where $a_{\rm frozen}=1/4$ and
$c_{\rm frozen}=3/4$ (see Fig.~9). These constants can be simply
understood.  Consider only the initial distribution of $C$ and $D$.
Regions between a pair successive $C$ (such regions are present with
probability 1/4) will be filled by $A$'s. Regions between a pair of $D$
as well as regions between a $C$ and a $D$ (present initially with
probability 3/4) will become $C$ domains.

\subsection{Symmetric Rule}

Let us now consider the $N$-species Lotka-Volterra model with
a {\it symmetric} eating rule, namely we assume that the $i^{\rm th}$
species can eat species $i-1\bmod N$ as well as $i+1\bmod N$. 

For $N=3$, all different species can eat each other without any
restriction. This model is thus equivalent to the 3-opinions voter
model (also called the stepping stone model).  In one dimension, the
concentration of interfaces is known to decay as $t^{-1/2}$, see {\it
e.g.}\cite{elp}.

For $N=4$, the situation is more interesting since {\it e.g.}  $A$ can
eat both $B$ and $D$ but cannot eat $C$.  Thus this model is different
from the 4-opinions voter model or the 4-species cyclic Lotka-Volterra
model. There are moving interfaces $M$ between
species $A$ and $B$, $B$ and $C$, $C$ and $D$, and $D$ and $A$, and
stationary interfaces $S$ between species $A$ and $C$ and species $B$
and $D$. Each moving interface is performing a random walk.  When a
moving interface meets a stationary one, the latter is eliminated,
$M+S\rightarrow M$; if two moving interfaces meet, they either produce
a stationary interface $M+M\rightarrow S$ or annihilate
$M+M\rightarrow \emptyset$ according to the state of the underlying
species.  On the mean-field level, this process is described by the rate
equations
\begin{equation}
\label{sm2}
\dot M=-4M^2,\quad
\dot S=M^2-SM.
\end{equation}
Eqs.~(\ref{sm2}), supplemented by the initial conditions $M(0)=1/2$
and $S(0)=1/4$, are solved to yield
\begin{equation}
\label{sm3}
M(t)={1\over 2+4t},\quad
S(t)={7\over 12}\,{1\over (1+2t)^{1/4}}
-{1\over 3+6t},
\end{equation}
implying the existence of two scales, $\ell\sim t^{1/4}$ and 
${\cal L}\sim t$. 

Fortunately, an exact analysis of the 4-species Lotka-Volterra model
with the symmetric eating rule is possible.  Moving interfaces do not
feel the stationary ones and they are undergoing diffusive
annihilation. As a result, their concentration decays according to
$M(t)\sim t^{-1/2}$.  Following the discussion in the previous
section, the fraction of stationary interfaces surviving from the
beginning is proportional asymptotically to the fraction of sites
which have not been visited by mobile interfaces up to time $t$,
$S(t)\sim P_0(t)\sim t^{-3/8}$ \cite{der}.  We should also take into
account creation of stationary interfaces by the annihilation of
moving interfaces. This process produces new stationary interfaces
with rate of the order $-dM/dt$ so the density of stationary
interfaces satisfies the rate equation
\begin{equation}
\label{stat0}
{dS\over dt}= {dP_0\over dt}-{dM\over dt}.
\end{equation}
Combining Eq.~(\ref{stat0}) with $P_0(t)\sim t^{-3/8}$ and $M(t)\sim
t^{-1/2}$, we find that interfaces which survive from the beginning
provide the dominant contribution while those created in the process
$M+M\rightarrow S$ contribute only to a correction of the order $t^{-1/8}$
\begin{equation}
\label{stat}
S(t)\sim t^{-3/8}\left[1+{\cal O}(t^{-1/8})\right].
\end{equation}
Thus a two-scale structure of the type (\ref{domains4}) emerges with typical
length , $\ell\sim t^{3/8}$ and ${\cal L}\sim
t^{1/2}$. The exponents for the 4-species Lotka-Volterra
with symmetric rules are summarized in Table 1. 
These asymptotic results agree only qualitatively
with the rate equations predictions.  Simulation results are in an
excellent agreement with these predictions, $M(t)\sim t^{-0.50}$ and
$S(t)\sim t^{-0.35}$ (see Fig.~10).  Refined analysis which makes use
of the expected correction of the order ${\cal O}(t^{-1/8})$ enables a
better estimate for the decay of stationary interfaces, namely
$S(t)\sim t^{-0.37}$.

\begin{figure}
\vspace{-.1in}
\narrowtext
\epsfxsize=\hsize
\epsfbox{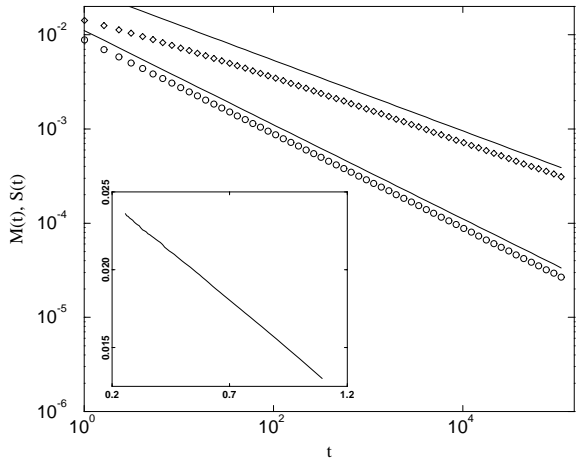}
\vspace{-.1in}
\caption{The concentrations of stationary (diamonds) and moving
(circles) interfaces as a function of time for the 4-species model
with a symmetric sequential dynamics. Lines of slope $1/2$ and $3/8$
are shown as references.  The insert shows $t^{3/8}S(t)$ as a function
of $t^{-1/8}$ where a straight line is expected.
\label{fig10}}
\end{figure}

The $N=5$ case with symmetric eating rules can be easily analyzed on
the level of the rate equations.  We omit the details as the analysis
is similar to the one presented in subsection III.D for the cyclic
model.  The conclusion is similar as well, namely the system
approaches a frozen state consisting of noninteracting domains.
Arguing as in the cyclic case we conclude that the threshold number of
species predicted by the mean-field rate equation approach is exact,
$N_c=5$, in agreement with our numerical simulations. 

We also found that the rate equation approach does not provide a
correct description of the decay of the mobile interfaces: $M_{\rm
MFT}(t)\propto e^{-t}$, while in the actual process $M(t)\propto
e^{-t^n}$ with $n$ close to $1/4$.  An upper bound, $n\leq 1/3$, can
be established by comparing to the trapping process, $M+T\to T$
\cite{bv}, where the survival probability $\sigma(t)$ for a particle
diffusing in a sea of immobile traps, $\sigma(t)\propto
\exp(-t^{1/3})$\cite{bv}, provides a lower bound for our original
problem, $M(t)\geq \sigma(t)$.

\subsection{Diffusion-Reaction Description: Cyclic Models}
 
So far we studied population dynamics occurring on a lattice.  Although
similar descriptions has been used in several other studies
\cite{t,ti,tome,may,sole}, the diffusion-reaction equation approach is
more popular \cite{lotka,montroll,murray}.  It is therefore useful to
establish a relationship between the two approaches.

To this end, consider a 3-species system with particles moving
diffusively and evolving according to the reaction scheme
(\ref{ABCscheme}), supplemented by reproduction and self-regulation.
On the level of a diffusion-reaction approach, this process is
described by the following partial differential equations
\begin{eqnarray}    
\label{ABC}
{\partial a\over \partial t}&=&
{\partial^2 a\over \partial x^2}+a(1-a)+ka(b-c), \nonumber\\
{\partial b\over \partial t}&=&
{\partial^2 b\over \partial x^2}+b(1-b)+kb(c-a), \\
{\partial c\over \partial t}&=&
{\partial^2 c\over \partial x^2}+c(1-c)+kc(a-b). \nonumber
\end{eqnarray}
In these equations, $a=a(x,t)$, $b=b(x,t)$, and $c=c(x,t)$ denote the
corresponding densities at point $x$ on the line; $a(1-a)$ is the Lotka
term describing reproduction and self-regulation; the diffusion constant
and the growth rates of each species are set equal to unity, and the 
constant $k$ measures the strength of the competition between species. 

For noninteracting species, $k=0$, and Eqs.~(\ref{ABC}) decouple to
the well-known single-species Fisher-Kolmogorov equations
\cite{murray,pattern}.  This equation has two stationary solutions,
$a=0$ and $a=1$; the former is unstable while the latter
is stable so any initial distribution approaches toward it.  Starting 
from an initial density close to stable equilibrium for $x<0$ and to
unstable equilibrium for $x>0$, a wave profile is formed and moves
into the unstable region \cite{murray,pattern,bram}.  The width of the
front is finite as a result of the competition between diffusion which
widens the front and nonlinearity which sharpens the front.

Consider now the case of interacting species, $k>0$.  The initial
dynamics is outside the scope of a theoretical treatment and should be
investigated {\it e.g.}  numerically solely on the basis of
Eqs.~(\ref{ABC}).  However, as the coarsening proceeds, single-species
domains form.  Inside say an $A$-domain, the density of $A$ species is
almost at stable equilibrium, $a(x,t)\cong 1$, while the densities of
$B$ and $C$ species are negligible.  In the boundary layer between say
an $A$ and a $B$ domains, the density of $C$ species is negligible.
Domain sizes grow while the width of boundary layer remains finite.
Therefore, in the long time limit one can treat boundary layers as
(sharp) interfaces which are expected to move into ``unstable''
domain.

To determine the velocity $v$ of the interface and the density
profiles we employ a well-known procedure \cite{pattern,bram}.
Consider an interface between say an $A$ domain to the left and a $B$
domain to the right.  We look for a wave-like solution, 
\begin{equation}
\label{wave}
a(x,t)=a(\xi), \quad
b(x,t)=b(\xi), \quad
\xi=x-vt.
\end{equation}
Substituting (\ref{wave}) into Eqs.~(\ref{ABC}) we arrive at a pair of
ordinary differential equations for the density profiles $a(\xi)$ and
$b(\xi)$.  These equations cannot be solved analytically.  To
determine the interface velocity, let us consider the densities far
 from the interface ($\xi=0$), say for $\xi\gg 1$.  In this region
$a(\xi)\ll 1$ and $b(\xi)\cong 1$.  Therefore the equation for $a(\xi)$
simplifies to
\begin{equation}
\label{appr}
a''+va'+(1+k)a=0,
\end{equation}
where $a'=da/d\xi$, {\it etc}.  By inserting an exponential solution,
$a(\xi)\sim e^{-\lambda \xi}$, into Eq.~(\ref{appr}) we get
$\lambda^2-v\lambda+(1+k)=0$.  In principle any velocity $v\geq
v_{\rm min}$, with $v_{\rm min}=2\sqrt{1+k}$, is possible.  This
resembles the situation with the Fisher-Kolmogorov
equation\cite{murray,pattern}.  According to 
the ``pattern selection principle'' \cite{murray,pattern},
 the minimum velocity is in fact realized for most initial
conditions.  The pattern selection principle is a {\it theorem} for
the Fisher-Kolmogorov equation (where the precise description of
necessary initial conditions is known)\cite{bram} while for many
other reaction-diffusion equations the pattern selection principle has
been verified numerically\cite{murray,pattern}.

Thus, for 3-species cyclic Lotka-Volterra model in 1D we have
established an asymptotic equivalence between the diffusion-reaction
approach and the lattice one with the parallel dynamics.  Given that
the density of interfaces decays as $t^{-1/2}$, one can anticipate the
same behavior for the diffusion-reaction model.  This result may be
difficult to observe directly from numerical integration of the
nonlinear partial differential equations (\ref{ABC}), and establishing
the complete relationship between lattice and diffusion-reaction
approaches remains a challenge for the future.

\subsection{Diffusion-Reaction Description: Symmetric Models}

Consider now the 3-species {\it symmetric} Lotka-Volterra on the level
of the diffusion-reaction description.  Rate equations of the type
Eqs.~(\ref{ABC}) are useless in this case since they do not contain
terms describing interactions among species.  Nevertheless it proves
useful to consider a similar symmetric system where interacting
species mutually annihilate upon collision.  The governing equations
read
\begin{eqnarray}    
\label{ABCsym}
{\partial a\over \partial t}&=&
{\partial^2 a\over \partial x^2}+a(1-a)-ka(b+c), \nonumber\\
{\partial b\over \partial t}&=&
{\partial^2 b\over \partial x^2}+b(1-b)-kb(c+a), \\
{\partial c\over \partial t}&=&
{\partial^2 c\over \partial x^2}+c(1-c)-kc(a+b). \nonumber
\end{eqnarray}

We again restrict ourselves to the late stages where a well-developed
domain structure has been already formed \cite{1species,2species}.  To
simplify the analysis further we assume that the competition is
strong, $k\to \infty$, so neighboring domains act as absorbing
boundaries.  We employ a quasistatic approximation, {\it i.e.} we
neglect time derivatives and perform a stationary analysis in a domain
of fixed size, and then make use of those results to determine the
(slow) motion of the interfaces.  Inside say an $A$ domain the density
$a(x)$ satisfies 
\begin{equation}    
\label{Asym}
{\partial^2 a\over \partial x^2}+a(1-a)=0.
\end{equation}
One should solve Eq.~(\ref{Asym}) on the interval $(0,L)$
subject to the boundary conditions $a(0)=a(L)=0$. 
The size of the domain, $L$, is assumed to be large compare to the width
of the interface, {\it i.e.}, $L\gg 1$.  In this limit, the flux of
$A$ species through the interface is equal to\cite{2species} 
\begin{equation}    
\label{flux}
F(L)\cong {1\over \sqrt{3}}-{\rm const.}\times e^{-L}.
\end{equation}
Clearly, if we have neighboring $L_1$-domain and $L_2$-domain,
then the smallest of the two domains shrinks while the largest grows, 
and the interface moves with velocity 
$F(L_1)-F(L_2)\propto e^{-L_2}-e^{-L_1}$.  Thus the average size
grows according to 
\begin{equation}    
\label{Lav}
{d\over dt} \langle L\rangle\propto \exp(-\langle L\rangle) 
\end{equation}
which is solved to yield $\langle L\rangle \sim \ln t$.  We see
that coarsening still takes place, but it is logarithmically slow.

Moreover, the determination of the complete domain size distribution
can be readily performed, at least numerically.  Clearly, in the late
stage all sizes are large, $L\gg 1$. Thus, only the smallest domain
shrinks and the two neighboring domains grow while other domains
hardly move at all.  This provides an extremal algorithm: (i) The
smallest domain $L_{\rm min}$ is identified; (ii) If the nearest
domains, $L_1$ and $L_2$, contain {\it similar} species, both
interfaces are removed and a domain of length $L_1+L_{\rm min}+L_2$ is
formed; (iii) If the nearest domains contain {\it dissimilar} species,
the two interfaces merge and form a new interface at the midpoint, and
thus domains of size $L_1+L_{\rm min}/2$ and $L_2+L_{\rm min}/2$ are
formed.  This process is identical to the 3-state Potts model with
extremal dynamics\cite{bdg}.  Similar one-dimensional models with
extremal dynamics have been investigated in a number of recent
studies\cite{kawasaki,dgy,bray,slava}.

Finally, we briefly discuss the symmetric rule model, with $N=2$ and
4.  In the two-species case \cite{1species,2species}, a mosaic of
alternating $A$ and $B$ domains is formed.  The late dynamics is again
extremal: the smallest domain merges with the two neighboring domains.
The domain-size distribution function\cite{kawasaki} and several
correlation functions\cite{bray} have been computed analytically.  In
the 4-species model, there are two independent kinetic processes: Slow
coarsening which takes place when domains of interacting species are
neighbors, and fast mixing of neighboring domains of non-interacting
species.  Clearly, almost all domains will be soon populated by pairs
of non-interacting species, so a mosaic of alternating $AC$ and $BD$
domains is formed, and the resulting dynamics of the symmetric
4-species model is {\it identical} to the dynamics of the 2-species
model.

Thus, in the symmetric case the reaction-diffusion approach provides
very different results compare to the lattice process.  It was also
seen that mapping onto a system of intervals with {\it extremal}
dynamics does provide an effective way to analyze the long
time behavior. 

\section{DISCUSSION}

In this paper, we investigated one-dimensional Lotka-Volterra systems
and found that they coarsen when the number of species is
sufficiently small, $N\leq 4$.  Typically, coarsening systems exhibit
dynamical scaling with a single scale\cite{br}.  When scaling holds,
analysis of the system is greatly simplified, {\it e.g.}, the single
scale grows as a power law, $\ell(t)\sim t^{\alpha}$, with
the exponent $\alpha$ independent of many details of the dynamics,
usually even independent of the spatial dimension\cite{br}.  In
contrast, for the Lotka-Volterra models we found that the coarsening
{\it depends} on details of the dynamics.  There are {\it two}
characteristic length scales: the average length of the single-species
domains, $\ell(t)\sim t^{\alpha}$, and the average length of
superdomains, ${\cal L}(t)\sim t^{\beta}$.  Precise definition of
superdomains depends on the number of species $N$: For $N=3$
interfaces between neighboring domains move ballistically and
superdomains are formed by strings of interfaces moving in the same
direction; for $N=4$, neighboring domains are typically
noninteracting, and superdomains are separated by active interfaces.

Dimensional analysis provides additional insight into the existence of
more than one scale.  Consider for simplicity parallel dynamics, where
the relevant parameters are the initial interface concentration $c_0$,
the interface velocity $v$, and time $t$.  There are only two
independent length scales, $c_0^{-1}$ and $vt$, and using dimensional
analysis one expects 
\begin{equation}    
\label{dim}
\ell(t)=vt\psi(c_0vt), \quad {\cal L}(t)=vt\Psi(c_0vt).
\end{equation}
If simple scaling holds, the length $c_0^{-1}$ set by initial
conditions should be irrelevant asymptotically.  Thus, the scaling
functions $\psi(z)$ and $\Psi(z)$ should approach constant values as
$z=c_0vt\to\infty$ implying $\ell(t)\sim {\cal L}(t)\sim vt$.  In
contrast, for the 3-species Lotka-Volterra model we found $\psi(z)\sim
z^{-1/2}$ when $z\to\infty$.  For the 4-species Lotka-Volterra model
both scaling functions exhibit asymptotic behavior different from the
naive scaling predictions, $\psi(z)\sim z^{-2/3}$ and $\Psi(z)\sim
z^{-1/3}$.  For the Lotka-Volterra model with symmetric eating rule
interfaces diffuse and thus the relevant length scales are $c_0^{-1}$
and $\sqrt{Dt}$.  Here, $\ell(t)=\sqrt{Dt}\psi(c_0^2 Dt)$ and ${\cal
L}(t)=\sqrt{Dt}\Psi(c_0^2 Dt)$.  When $N=4$, the two scale structure
implies $\psi(z)\sim z^{-1/8}$ as $z\to\infty$.

Thus simple dynamical scaling is violated for the one-dimensional
Lotka-Volterra models.  Violations of scaling have been reported in a
few recent studies of coarsening in one- and two-dimensional
systems\cite{cz,bh,nbm,mg,rb,r,zgg,zz}.  To the best of our knowledge,
however, in previous work violations of dynamical scaling have been
seen only in some systems with vector and more complex order
parameter.  In contrast, Lotka-Volterra models can be interpreted as
systems with {\it scalar} order parameter, although the number of
equilibrium states $N$ generally exceeds two, the characteristic
value for Ising-type systems.

Finally we note that presence of only two length scales exemplifies
the mildest violation of classical single-size scaling.  Generally, if
scaling is violated one expects the appearance of an infinite number
of independent scales, {\it i.e.}, multiscaling\cite{cz,cz1}.
Surprisingly, we found no evidence of multiscaling.  Similar
two-length scaling has been observed in the simplest one-dimensional
system with vector order parameter, namely in the XY model\cite{rb},
and in the single-species annihilation with combined diffusive and
convective transport\cite{brk}.  Indications of the three-length
dynamical scaling have been reported in the context of
coarsening\cite{zgg} and chemical kinetics\cite{lr,redner}.

\section{SUMMARY}

In this study, we addressed the dynamics of competitive immobile
species forming a cyclic food chain. We first examined a cyclic model
with asymmetric rules and symmetric initial conditions and have
observed a drastic difference between the two extremes, corresponding
to the complete graph (``infinite-dimensional'') and to
one-dimensional substrates.  In the latter case, spatial
inhomogeneities develop, and the resulting kinetic behavior is very
sensitive to the number of species.  For a sufficiently small number
of species, the system coarsens and is described by a set of exponents
summarized in Table 1.  These exponents {\it depend} on the number of
species and on the type of dynamics (parallel or sequential).  Thus,
to describe coarsening in systems with {\it non-conservative} dynamics 
it is necessary to specify the details of the dynamics.

The time distribution of the number of mutations has also been
investigated and we presented scaling arguments as well as an exact
result for a particular case.  We also treated symmetric interaction
rules.  This system is especially interesting when $N=4$ as it
provides a clear realization of the recently introduced notion of
``persistent'' spins in terms of the stationary interfaces.  Finally, 
we discussed a relationship to the alternative reaction-diffusion
equations description.  While for the cyclic version both the lattice
and the reaction-diffusion approaches have been found to be closely
related, for the symmetric version very different results have emerged
and a relationship with extremal dynamics has been established.

\vskip 0.2in

We thank S.~Ispolatov, G.~Mazenko, J.~Percus, and
S.~Redner for discussions.  L.~F. was supported by the Swiss NSF,
P.~L.~K. was supported in part by a grant from NSF, 
E.~B.~ was supported in part by NSF Award Number 92-08527, and by
the MRSEC Program of the NSF under Award Number DMR-9400379.

\end{multicols} 
\end{document}